\documentclass[aip,jcp,twocolumn,showpacs,nobibnotes,reprint]{revtex4-1}
\usepackage{graphics,graphicx,amsfonts,amsmath,amsbsy,amssymb,color}
\usepackage{verbatim}  
\usepackage{psfrag}  
\usepackage{tabularx}
\usepackage{xmpmulti}
\usepackage{hyperref}
\usepackage{marginnote}
\usepackage{subfigure}
\usepackage{paracol}
\usepackage{xparse}
\usepackage{layouts}
\usepackage{afterpage}
\usepackage{braket}
\usepackage{paralist}
\usepackage{bm}
\usepackage{url}
\usepackage[normalem]{ulem}
\usepackage{color}
\usepackage[colorinlistoftodos]{todonotes}
\usepackage[american]{babel} 

\newcommand{\refeq}[1]{{Eq.~(\ref{#1})}}
\newcommand{\reffig}[1]{{Fig.~\ref{#1}}}

\hypersetup{hyperindex,breaklinks,colorlinks=true,linkcolor=blue,citecolor=blue,urlcolor=blue,filecolor=blue}

\begin{document}

\title{An optimized twist angle to find the twist-averaged correlation energy applied to the uniform electron gas}

\author{Tina~Mihm}
\address{Department of Chemistry, University of Iowa}
\address{University of Iowa Informatics Initiative, University of Iowa}

\author{Alexandra~R.~McIsaac}
\address{Department of Chemistry, Massachusetts Institute of Technology}

\author{James~J.~Shepherd}
\email{james-shepherd@uiowa.edu}
\address{Department of Chemistry, University of Iowa}
\address{University of Iowa Informatics Initiative, University of Iowa}

\pacs{71.10.Ca, 71.15.Ap}
\begin{abstract}
We explore an alternative to twist averaging in order to obtain more cost-effective and accurate extrapolations to the thermodynamic limit (TDL) for coupled cluster doubles (CCD) calculations.
We seek a single twist angle to perform calculations at, instead of integrating over many random points or a grid. 
We introduce the concept of connectivity, a quantity derived from the non-zero four-index integrals in an MP2 calculation. 
This allows us to find a special twist angle that provides appropriate connectivity in the energy equation, and which yields results comparable to full twist averaging.
This special twist angle effectively makes the finite electron number CCD calculation represent the TDL more accurately, reducing the cost of twist-averaged CCD over $N_\mathrm{s}$ twist angles from $N_s$ CCD calculations to $N_s$ MP2 calculations plus one CCD calculation. 
\end{abstract}
\date{\today}
\maketitle

\section{Introduction}

In recent years, the use of wavefunction-based post-Kohn--Sham or post-Hartree--Fock methods to solve problems in materials science has proliferated. \cite{muller_wavefunction-based_2012} 
This is in part driven by an interest in obtaining precise energies (accurate to within 1mHa) for complex systems using hierarchies of methods found in quantum chemistry such as coupled cluster theory.
While growing in popularity, wavefunction methods have yet to see widespread adoption, in large part due to their significant computational cost scaling with system size.
This is especially of note in coupled cluster theory using a plane wave basis, and as a result, some authors are seeking methods to control finite size errors in order to run calculations using smaller system sizes.~\cite{gruber_applying_2018}

Finite size errors arise when attempts are made to simulate an infinite system Hamiltonian with a periodic supercell containing a necessarily finite particle number.\cite{fraser_finite-size_1996,drummond_finite-size_2008} 
The finite size of a supercell places a limitation on the minimum momenta in Fourier sums (e.g., with a cubic box of length $L$, the smallest momentum transfer is $2\pi/L$).
These limitations ultimately lead to errors in the correlation energy;  \cite{gruber_applying_2018,ruggeri_correlation_2018} 
this has been attributed to  long range van der Waals forces. \cite{gruber_applying_2018,gruber_ab_2018}

Since these finite size errors are large and slowly converging with increasing supercell size, {which has been analyzed in detail for coupled cluster theory, \cite{mcclain_gaussian-based_2017}} there has been significant interest in developing wavefunction methods with reduced computational cost to circumvent finite size error and allow the treatment of larger supercells.
These include embedding methods,\cite{sun_quantum_2016} 
such as density matrix embedding,\cite{knizia_density_2012,knizia_density_2013,bulik_electron_2014,bulik_density_2014,ricke_performance_2017,zheng_cluster_2017,pham_can_2018} 
wavefunction-in-DFT embedding,\cite{henderson_embedding_2006,tuma_treating_2006,gomes_calculation_2008,sharifzadeh_all-electron_2009,huang_quantum_2011,libisch_embedded_2014, manby_simple_2012,goodpaster_accurate_2014,chulhai_projection-based_2018} 
electrostatic embedding,\cite{hirata_fast_2005,dahlke_electrostatically_2007,hirata_fast_2008,leverentz_electrostatically_2009,bygrave_embedded_2012} 
QM/MM-inspired schemes,\cite{shoemaker_simomm:_1999,sherwood_quasi:_2003,herschend_combined_2004,beran_predicting_2010,chung_oniom_2015}
and others.\cite{eskridge_local_2018,lan_communication:_2015,rusakov_self-energy_2019,voloshina_embedding_2007,masur_fragment-based_2016}
Local correlation methods\cite{collins_energy-based_2015,usvyat_periodic_2018}
such as fragment-based schemes,\cite{gordon_fragmentation_2012,li_generalized_2007,li_cluster--molecule_2016,rolik_general-order_2011,li_divide-and-conquer_2004,kobayashi_alternative_2007,kristensen_locality_2011,ghosh_noncovalent_2010,kitaura_fragment_1999,fedorov_extending_2007,netzloff_ab_2007,ziolkowski_linear_2010}
incremental methods,\cite{stoll_correlation_1992,paulus_method_2006,friedrich_fully_2007,stoll_approaching_2012,friedrich_incremental_2013,voloshina_first_2014,kallay_linear-scaling_2015,fertitta_towards_2018}
and heirarchical methods,\cite{deev_approximate_2005,manby_extension_2006,nolan_calculation_2009,collins_ab_2011}
break the system into smaller subsystems, then extrapolate or stitch together the energies.
Some methods take advantage of range separation\cite{toulouse_adiabatic-connection_2009,bruneval_range-separated_2012,shepherd_range-separated_2014}
or other distance-based schemes\cite{spencer_efficient_2008,maurer_efficient_2013,kats_sparse_2013,kats_speeding_2016,ayala_extrapolating_1999}
to reduce computational cost.  

In addition to work on developing or modifying electronic structure methods, much work on reducing the cost of wavefunction methods has been focused on modifying basis sets in order to accelerate convergence and decrease computation time.
Local orbital methods have been popular,\cite{pisani_local-mp2_2005,ayala_atomic_2001,usvyat_periodic_2015,werner_fast_2003,flocke_natural_2004,werner_scalable_2015,rolik_efficient_2013,forner_coupled-cluster_1985,schutz_low-order_2000,neese_efficient_2009,sun_gaussian_2017,booth_plane_2016,blum_ab_2009,subotnik_local_2005}
often based on the local ansatz of Pulay and Saebo\cite{saebo_local_1993} or Stollhoff and Fulde.\cite{stollhoff_local_1977}
Other common methods include progressive downsampling,\cite{shimazaki_brillouin-zone_2009,hirata_fast_2009,ohnishi_logarithm_2010}
downfolding,\cite{purwanto_frozen-orbital_2013} 
use of explicitly-correlated basis sets\cite{adler_local_2009,shiozaki_communications:_2010,gruneis_explicitly_2013,usvyat_linear-scaling_2013,gruneis_efficient_2015}
or natural orbitals,\cite{gruneis_natural_2011} 
and tensor manipulations.\cite{hohenstein_tensor_2012,benedikt_tensor_2013,hummel_low_2017,peng_highly_2017,motta_efficient_2018}
Discussion of the details and relative merits of these methods is beyond the scope of this paper; for a review, we direct the interested reader to Refs. \onlinecite{huang_advances_2008,muller_wavefunction-based_2012,beran_modeling_2016,andreoni_coupled_2018}.

However, there has been some work on developing corrections for finite size errors.\cite{fraser_finite-size_1996,kent_finite-size_1999,kwee_finite-size_2008,drummond_finite-size_2008,holzmann_theory_2016,liao_communication:_2016}
Many-body methods can sometimes be integrated to the thermodynamic limit (TDL),\cite{gell-mann_correlation_1957,nozieres_correlation_1958,onsager_integrals_1966,bishop_electron_1982,bishop_electron_1978,bishop_overview_1991,ziesche_selfenergy_2007}
allowing for the derivation of analytic finite-size correction expressions.\cite{chiesa_finite-size_2006} 
Several studies from the last year have particular relevance to our work here. Gr{\"u}eneis \emph{et al.}\cite{gruber_applying_2018,gruber_ab_2018} employed a grid integration within periodic coupled cluster for \emph{ab initio} Hamiltonians with applications to various solids.
In another study, Alavi \emph{et al.}\cite{ruggeri_correlation_2018} devised a novel extrapolation relationship that links different electron gas calculations through the density parameter. 
Both of these papers use a technique known as twist averaging to try to remove finite size error.

Twist averaging is a method that attempts to control finite size errors by first offsetting the $k$-point grid by a small amount, ${\bf k}_s$, and then averaging over all possible offsets.\cite{lin_twist-averaged_2001}
{
We refer to ${\bf k}_s$ here as a twist angle.
One of the main purposes of twist averaging is to provide for a smoother extrapolation to the thermodynamic limit by reducing severe energy fluctuations as the particle number varies.
}
When performed with a fixed particle number and box length, this process is referred to as twist averaging in the canonical ensemble, which is what we study here. 
When employed in stochastic methods, such as variational Monte Carlo,\cite{lin_twist-averaged_2001} diffusion Monte Carlo\cite{drummond_finite-size_2008} or full configuration interaction quantum Monte Carlo,\cite{ruggeri_correlation_2018,shepherd_quantum_2013} the grid can be stochastically sampled at the same time as the main stochastic algorithm, and both stochastic error and error in twist-averaging related to approximate integration can be removed at the same time. 
As a result, the scaling with the number of twist angles sampled is extremely modest. 
Unfortunately, the same cost savings cannot be realized for deterministic methods. 
In this case, in order to achieve a reasonable estimate for the average, one must use a large number of individual energy calculations. 
This results in the cost scaling linearly with the number of twist angles used, {although the lessening of finite size effects with rising electron number would alleviate this scaling to some extent.\cite{mcclain_gaussian-based_2017}}

Here, we seek to remedy the linear scaling of twist averaging for deterministic methods by devising a way to provide an energy that is as accurate as twist-averaging, but with single-calculation cost. 
{\color{black}
In principle, it is possible to find a single twist angle which exactly reproduces the total twist-averaged energy by recognizing that it is an integral of the energy over the twist angles for a system.
This was the same logic used in analysis by Baldereschi to find a special $k$-point\cite{baldereschi_mean-value_1973} and has been used by others in the QMC community to find a special twist angle.\cite{dagrada_exact_2016,Rajagopal_quantum_1994,Rajagopal_variational_1995}
We are motivated similarly and wish to find a single twist angle that yields an energy approximately equal to the full twist-averaged energy for CCD and related wavefunction methods.}
{
We take advantage of the similarity between the MP2 and CCD correlation energy expressions, using the much cheaper MP2 method to find a single twist angle that produces a system with the most similar number of allowed excitations to the twist-averaged system.
We refer to this set of allowed excitations as the `connectivity'. 
We then use this twist angle to calculate the CCD energy, which is in good agreement with the fully twist-averaged CCD energy.
Finally, we compare our energies to those obtained using one twist angle at the Baldereschi point.\cite{baldereschi_mean-value_1973} }
{ We do not seek to completely remedy the whole of the finite size error, instead noting that other authors have come up with corrections or extrapolations that can be used after twist-averaging is applied.\cite{chiesa_finite-size_2006,drummond_quantum_2009,gruber_ab_2018}}

\section{Twist averaging \& Connectivity}

\begin{figure}
\includegraphics[width=0.49\textwidth,height=\textheight,keepaspectratio]{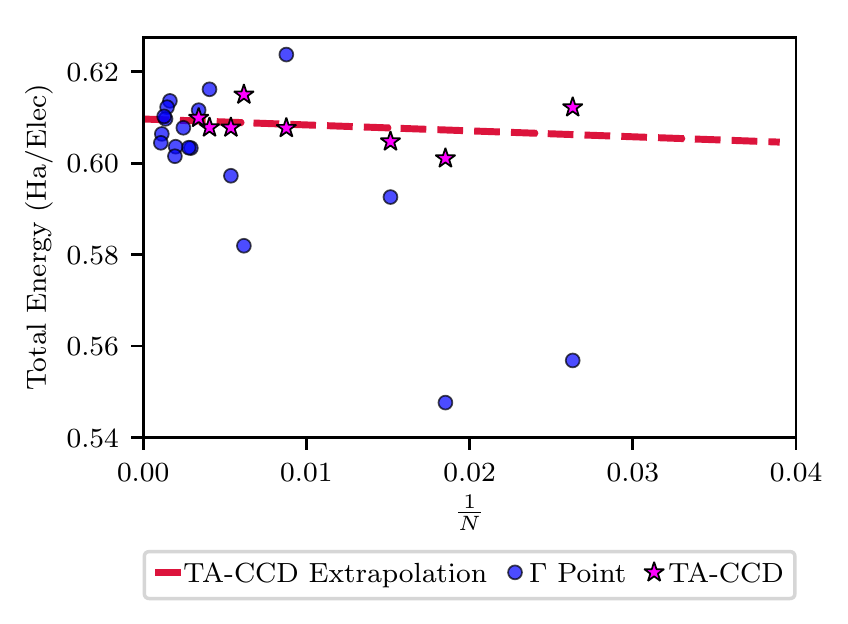}
\caption{Comparison between the twist-averaged (TA) CCD energy and the $\Gamma$-point CCD energy for a uniform electron gas with $r_s=1.0$ as the system size changes (up to $N=294$ and $N=922$, respectively). In general, an extrapolation (here, red line) is performed to calculate the TDL energy. Twist averaging makes this extrapolation easier, because the noise around the extrapolation is smaller, leading to a smaller extrapolation error. Twist averaging is performed over 100 twist angles. Standard errors are calculated in the normal fashion for twist averaging, $\sigma\approx\sqrt{\mathrm{Var}(E_{\mathrm{CCD}}({{\bf k}_s})) / N_s}$ (are too small to be shown on the graph, on average 0.2 mHa/el).}

\label{fig:TADemonstration}
\end{figure}

Both continuum/real-space and basis-set twist averaging have been used effectively in quantum Monte Carlo calculations;\cite{lin_twist-averaged_2001,drummond_finite-size_2008,ruggeri_correlation_2018,shepherd_quantum_2013} however, twist averaging remains relatively rare in coupled cluster calculations. 
In \reffig{fig:TADemonstration}, the total $\Gamma$-point CCD energy ($N=38$ to $N=922$) and twist-averaged CCD energy ($N=38$ to $N=294$) are plotted alongside the extrapolation to the TDL for the uniform electron gas ($0.609(3)$ Ha/electron, where the error in the last digit is in parentheses). The CCD calculation is performed in a finite basis that is analogous to a minimal basis.\cite{shepherd_many-body_2013} The $\Gamma$-point energy is highly non-monotonic; it does not fit well with the extrapolation. The twist-averaged data shows a much better fit with the extrapolation, resulting in a better estimate of the TDL. The drawback of twist averaging, however, is that it costs $N_\mathrm{s}\,\mathcal{O}\mathrm{[CCD]}$ for $N_\mathrm{s}$ twist angles (here, 100). The twist-averaged energy becomes too costly to calculate with CCD for system sizes above 294 electrons.

Figure \ref{fig:TADemonstration} is a clear statement of the problem we wish to resolve here. 
{ Twist averaging resolves some finite size errors that are present at an individual particle number $N$, and allows for improved extrapolation to the thermodynamic limit. }
That said, the scaling with the number of twist angles is cost-prohibitive. We aim to develop an approximation to twist averaging that gives comparable accuracy at a fraction of the cost. We begin by analyzing how the Hartree-Fock energy and the MP2 correlation energy are modified by twist averaging.
This analysis then allows us to build an algorithm that produces CCD twist-averaged accuracy/results for only MP2 cost. 

\subsection{Hartree-Fock and single-particle eigenvalues} 

{
\begin{figure}
\includegraphics[width=0.49\textwidth,height=\textheight,keepaspectratio]{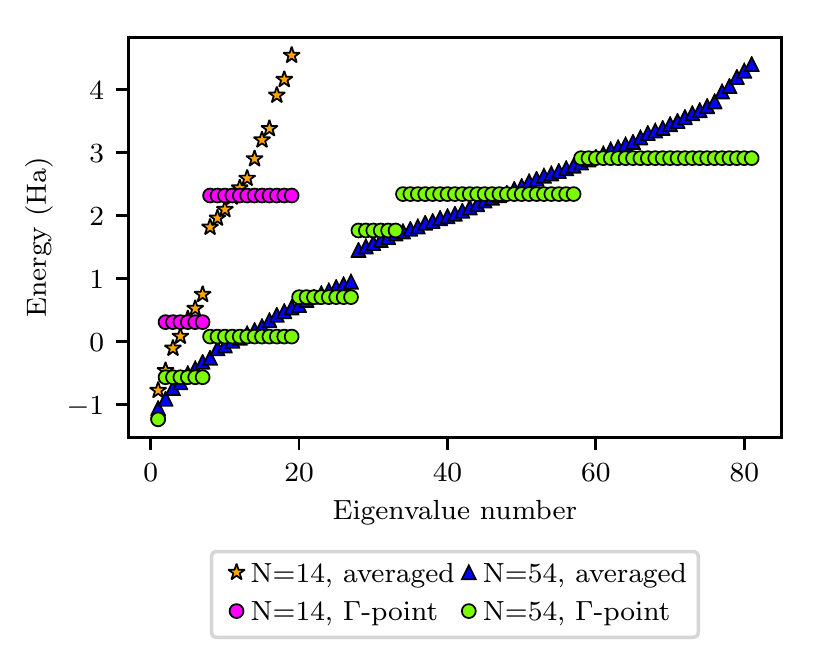}
\caption{{The degeneracy pattern in the energy levels of the $\Gamma$-point calculation can be identified by plotting the HF eigenvalues are plotted in ascending order. Here, we show $N=14, 54$ two systems that are closed shell at the $\Gamma$-point. Averaging the eigenvalues in the manner described in the text removes these degeneracies. The gap between the eigenvalues themselves and across the band gap goes to zero as the TDL is approached, giving rise to the metallic character of the gas.}}
\label{fig:AverageEigenvalues}
\end{figure}
}

A finite-sized electron gas at the $\Gamma$-point is only closed-shell at certain so-called magic numbers, which are determined by the symmetry of the lattice (for example $N=2$, 14, 38, and 54).
{ One of the reasons that the $\Gamma$-point calculations are so noisy (\reffig{fig:TADemonstration}) is that there are degeneracies in the HF eigenvalues, which can be seen in \reffig{fig:AverageEigenvalues} and has long been recognized. \cite{drummond_quantum_2009} This can be partially remedied by modifying the Hartree--Fock eigenvalues. The starting-point for this is} writing the HF energy as follows:
\begin{equation}
E_\mathrm{HF}({\bf k}_s)= \sum_{i} T_i ({\bf k}_s)- \frac{1}{2} \sum_{ij} v_{ijji}({\bf k}_s)
\label{eq:fullHF}
\end{equation}
where $T_i$ is the kinetic energy of orbital $i$ and  $v_{ijji}$ is the exchange integral between electrons in orbitals $i$ and $j$.
Here, we have included the explicit form of the dependence on the twist angle, ${\bf k}_s$.

The twist-averaged energy is found by summing \refeq{eq:fullHF} over all possible ${\bf k}_s$:
\begin{equation}
\langle E_\mathrm{HF} \rangle_{\bf{k}_s} = 
\frac{1}{N_\mathrm{s}}\sum_{{\bf k}_s}^{N_\mathrm{s}} \sum_{i} T_i ({\bf k}_s) - \frac{1}{N_\mathrm{s}}\sum_{{\bf k}_s}^{N_\mathrm{s}}\frac{1}{2} \sum_{ij} v_{ijji}({\bf k}_s) 
\end{equation}
where $N_s$ indicates the number of twist angles used. Swapping the sums yields:
\begin{equation}
\langle E_\mathrm{HF} \rangle_{\bf{k}_s} = \sum_{i} \left[\frac{1}{N_\mathrm{s}} \sum_{{\bf k}_s}^{N_\mathrm{s}}  T_i ({\bf k}_s) \right]- \frac{1}{2} \sum_{ij} \left[ \frac{1}{N_\mathrm{s}}\sum_{{\bf k}_s}^{N_\mathrm{s}} v_{ijji}({\bf k}_s) \right] .
\label{eq:ks_sums}
\end{equation}
Therefore, twist averaging the HF energy is numerically identical to twist averaging the individual matrix elements:
\begin{equation}
\langle E_\mathrm{HF} \rangle_{\bf{k}_s} = \sum_{i}  \langle T_i \rangle_{\bf{k}_s} 
- \frac{1}{2} \sum_{ij} \langle v_{ijji} \rangle_{\bf{k}_s}  
\end{equation}
Overall, then, we can use twist-averaged HF eigenvalues in place of twist-averaging the HF energy, obtaining a more reasonable density of states \reffig{fig:AverageEigenvalues}.
We will use this in our subsequent scheme.

\subsection{Beyond Hartree--Fock} 

{ The above approach does not generalize to correlated theories because they have more complex energy expressions. For example, averaging the second-order M{\o}ller-Plesset theory (MP2) correlation energy over all possible twist angles can be written:
\begin{equation}
\langle E_\mathrm{corr} \rangle_{\bf{k}_s}=\frac{1}{N_\mathrm{s}}\sum_{{\bf k}_s}^{N_\mathrm{s}} \frac{1}{4}\sum_{ijab} \bar{t}_{ijab}({\bf k}_s) \bar{v}_{ijab} ({\bf k}_s),
\label{eq:Mp2}
\end{equation}
where $i$ and $j$ refer to occupied orbitals and $a$ and $b$ refer to unoccupied orbitals. The symbols $\bar{v}$ and $\bar{t}$ refer to the antisymmetrized electron-repulsion integral and amplitude respectively. 
For MP2:
\begin{equation}
\bar{t}_{ijab} ({\bf k}_s) \bar{v}_{ijab} ({\bf k}_s)=
\frac{ |\bar{v}_{ijab}({\bf k}_s)|^2 }{\epsilon_i({\bf k}_s)+\epsilon_j({\bf k}_s)-\epsilon_a({\bf k}_s)-\epsilon_b({\bf k}_s)}
\end{equation}}

{ Even though MP2 diverges in the thermodynamic limit, the energy expression (\refeq{eq:Mp2}) has a similar structure to coupled cluster theory, the random phase approximation, and even full configuration interaction quantum Monte Carlo. As such, we can make generalized observations using the MP2 energy expression, and then use these observations to derive a scheme to find an optimal $k_s$ twist angle that works for all of these methods.}

\subsection{The connectivity approach}

The MP2 correlation energy can vary substantially as the twist angle is changed. For example, in the $N=14$ electron system with a basis set of $M=38$ orbitals, the MP2 energy can vary between $-0.0171$ Ha/electron to $-0.0001$ Ha/electron. 
This arises, in particular, because the number of low-momentum excitations (minimum $|{\bf k}_i-{\bf k}_a|$) will vary significantly.
Since the contribution of each excitation to the MP2 sum is $|{\bf k}_i-{\bf k}_a|^{-4}$, there is a rapid decay of an excitation's contribution to the correlation energy beyond the minimum vector. 

{ This effect arises because, when the twist angle is changed,} different orbitals now fall into the occupied ($ij$) space, and different orbitals fall into the virtual ($ab$) space. This changes the value of the sum over both occupied and virtual orbitals, since many individual terms in the sum are now substantively different. We illustrate this using a diagram in the Supplementary Information.

By contrast, the integrals themselves do not change; { to show this, the integral can be written:
{\color{black}
\begin{equation}
v_{ijab}=\frac{4\pi}{L^3} \frac{1}{({\bf k}_i-{\bf k}_a)^2} \delta_{{\bf k}_i-{\bf k}_a , {\bf k}_b-{\bf k}_j} \delta_{\sigma_i \sigma_a}\delta_{\sigma_j \sigma_b} .
\label{eq:ERIs}
\end{equation}
}
The Kronecker deltas, $\delta$, ensure that momentum and spin symmetry (denoted $\sigma$) are conserved.}
 On changing ${\bf k}_p \rightarrow {\bf k}_p+{\bf k}_s$ for all ${\bf k}$'s, the difference in the denominator here does not change, since $({\bf k}_i+{\bf k}_s-{\bf k}_a-{\bf k}_s)^2=({\bf k}_i-{\bf k}_a)^2$. {\color{black} In general, our calculations were set up using details which can be found in our prior work e.g. Ref. \onlinecite{shepherd_convergence_2012}}.

At this stage, we conjecture that \emph{if} one of the mechanisms by which twist averaging is affecting the MP2 energy (and other correlation energies) is to smooth out the inconsistent contributions between different momenta, \emph{then} it might be possible for us to find a `special twist angle' where the number of low-momentum states for that single twist angle is a good match to the average number of momentum states across all twist angles. { Further, we will show this special twist angle is transferable to other, more sophisticated methods such as coupled cluster doubles theory. }

To find this special twist angle, we proceed as follows:
\begin{enumerate}
\item For a given twist angle ${\bf k}_s$, loop over the same $ijab$ as the MP2 sum $\sum_{ijab}$. For each $ijab$ set:
\begin{enumerate}
\item Determine the momentum transfer $x=|{\bf n}_i-{\bf n}_a|^2$ where ${\bf n}_a$ is the integer equivalent of the quantum number: ${\bf k}_a=\frac{2\pi}{L}{\bf n}_a$.
\item Increment a histogram element $h_x$ by one. 
\end{enumerate}
\item Create a vector ${\bf h}$, whose elements are $h_x$, which correspond to the number of of $v_{ijab}$ matrix elements with magnitude $\frac{1}{\pi L}\frac{1}{x}$ that are encountered during the MP2 sum. 
\item Average ${\bf h}$ over all twist angles, yielding $\langle {\bf h} \rangle_{\bf{k}_s}$
\item Loop over the twist angles again, and find the single ${\bf h}$ (and corresponding twist angle) that best matches $\langle {\bf h} \rangle_{\bf{k}_s}$ using:
\begin{equation}
\min_{\bf{k}_s}  \sum_x \frac{1}{x^2} \left( h_x - \langle h_x \rangle_{\bf{k}_s} \right)^2
\end{equation}
The weight term $1/x^2$ was chosen empirically to diminish the contributions of large numbers of high-momentum weights that contribute relatively little to the energy.
\end{enumerate}

{
Looking at \refeq{eq:Mp2}, there are two ways to proceed. We could either use this special ${\bf k}_s$ for all aspects of the calculation (e.g. for both the integral evaluation and the eigenvalue difference), or we could use the special ${\bf k}_s$ for the integral only, and twist-average the eigenvalues before performing the CCD calculation. 
We found that the latter was more numerically effective for $N=14$ and decided to use this approach to generate the results presented here. 
In general, though, for larger systems it does not make a large difference.}

In practice, we implemented this algorithm within an MP2 and CCD code; we call the MP2 calculation at each twist angle and then the CCD calculation once at the end. 
For the remainder of this work, we will call this application of the above algorithm the ``connectivity scheme," referencing the idea that the pattern of non-zero matrix elements $v_{ijab}$ resembles a connected network.

\section{Results}

\begin{figure}
\begin{center}
\vspace{-1cm}
\subfigure[\mbox{}]{%
\includegraphics[width=0.4\textwidth,height=\textheight,keepaspectratio]{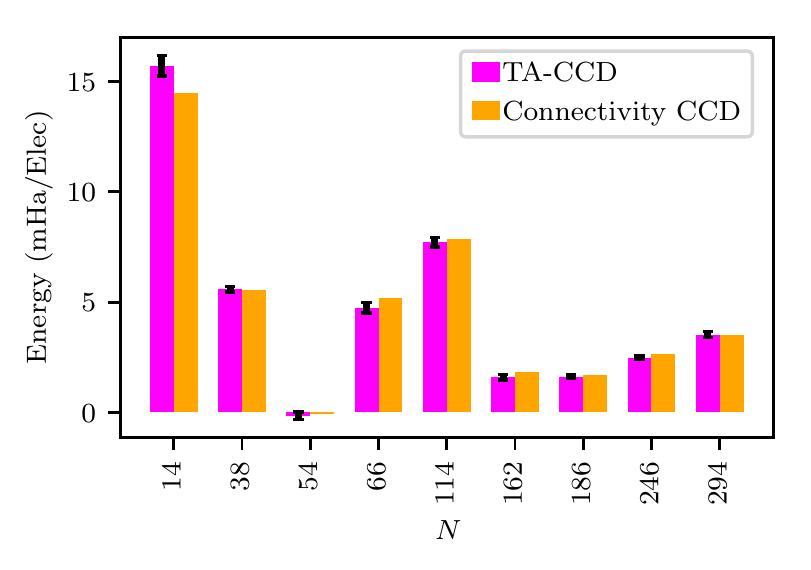}
\label{subfig:diffN}
}

\subfigure[\mbox{}]{%
\includegraphics[width=0.4\textwidth,height=\textheight,keepaspectratio]{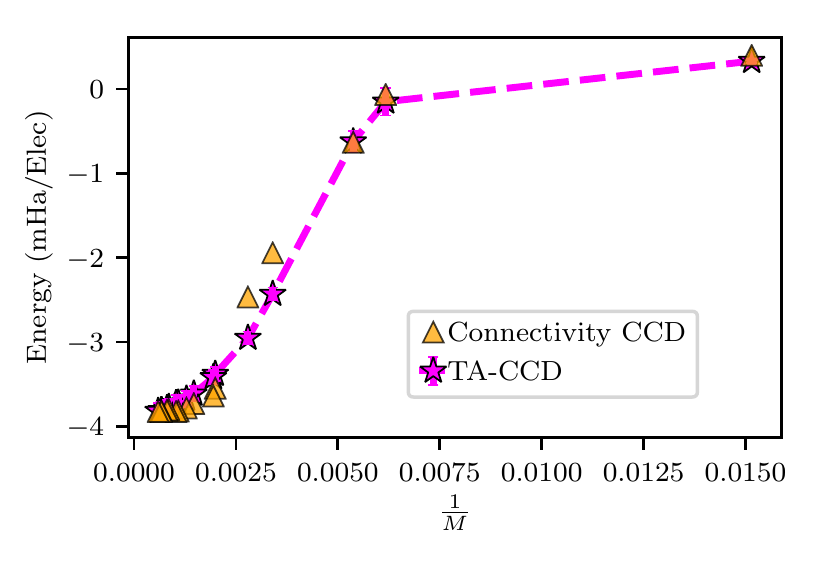}
\label{subfig:diffM}
}

\subfigure[\mbox{}]{%
\includegraphics[width=0.4\textwidth,height=\textheight,keepaspectratio]{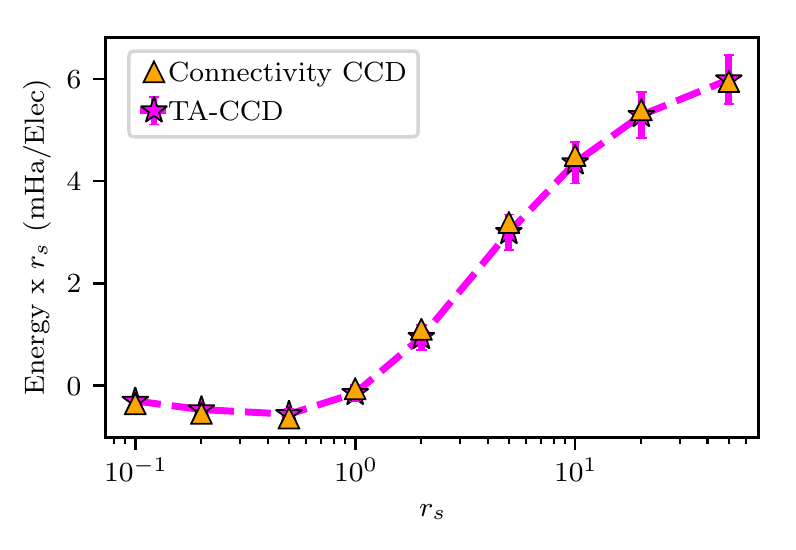}
\label{subfig:diffRs}
}

\caption{All energies shown represent the difference in correlation energy between the $\Gamma$-point and the relevant calculation, since, by design, the Hartree-Fock energy is identical between the connectivity scheme and standard twist averaging (TA).  The connectivity scheme delivers comparable corrections to the correlation energy (relative to the $\Gamma$-point) when compared with twist averaging across a wide range of (a) electron numbers (using a minimal basis set, where $M \approx 2N$, as mentioned in Ref. \onlinecite{shepherd_many-body_2013} and tabulated in the Supplementary Information), (b) different basis sets ($M=36-2838$ orbitals, with $N=54$ electrons), and (c) $r_s$ values (0.01 -- 50.0 a.u., with $N=54$ electrons). Twist averaging is performed over 100 twist angles. Standard errors are calculated in the normal fashion for twist averaging, $\sigma\approx\sqrt{\mathrm{Var}(E_{\mathrm{CCD}}({{\bf k}_s})) / N_s}$. }\label{fig:results}
\end{center}
\end{figure}

{We demonstrate the effectiveness of this algorithm for coupled cluster calculations on the uniform electron gas in \reffig{fig:results}. 
In general, our results as show that the connectivity scheme works for different electron numbers, basis sets, and $r_s$ values.
Furthermore, evaluation of the connectivity scheme is approximately 100x cheaper than twist averaging. }

In \reffig{subfig:diffN}, we compare the connectivity scheme to full twist-averaging for CCD calculations on the uniform electron gas. Energy differences from the $\Gamma$-point energy are plotted for each electron number. Our results show that the connectivity scheme delivers comparable accuracy (mean absolute deviation = 0.3 mHa/electron) to twist averaging, with the benefit of being much faster to compute. The connectivity scheme is substantially cheaper than the twist-averaging scheme: the $N=294$ twist-averaged calculation, for example, costs $58$ hours, which is about the same time it takes to run the $N=922$ connectivity scheme calculation. A complete set of timings is provided in the Supplementary Information. 

In \reffig{subfig:diffM}, we compare our connectivity scheme to full twist-averaging over a range of basis set sizes ($M= 36 - 2838$ orbitals) for 54 electrons. In \reffig{subfig:diffRs}, we compare the connectivity scheme to full twist-averaging over a range of $r_s$ values ($0.01 -50.0$ a.u.) for 54 electrons.  In both cases there is good agreement between the two methods for all system sizes, proving that the connectivity scheme delivers good accuracy when compared with twist averaging for a range of both basis set sizes (mean absolute deviation $<$ 0.35 mHa/electron) and $r_s$ values (mean absolute deviation $<$ 0.25 mHa/electron) at a decreased cost.

\begin{figure}
\includegraphics[width=0.5\textwidth,height=\textheight,keepaspectratio]{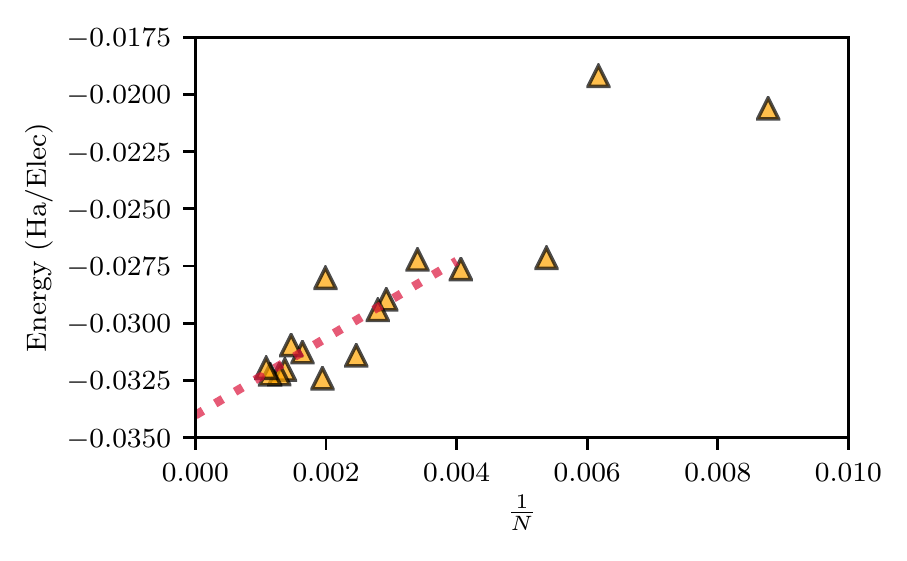}
\label{subfig:TDL1}

\caption{Connectivity scheme CCD correlation energies for electron numbers up to $N=922$ for $r_s=1.0$ in the uniform electron gas (yellow triangles). We fit 10 points (dotted red line) to the function $E=a+bN^{-1}$, as proposed by other authors; \cite{drummond_finite-size_2008} we then use this fit to extrapolate to the thermodynamic limit.}
\label{fig:TDLresults}
\end{figure}

In \reffig{fig:TDLresults}, we show the extrapolation of our connectivity scheme CCD correlation energy to the thermodynamic limit for the $r_s=1.0$ uniform electron gas. We perform calculations up to $N=922$ electrons, and fit these results to the equation $E=a+bN^{-1}$, as proposed by other authors. \cite{drummond_finite-size_2008}  We then use this fit to extrapolate the correlation energy to the thermodynamic limit. We also performed the same extrapolation for the twist-averaged data set up to $N=294$ electrons (not shown). The extrapolations predict the TDL energy to be $-0.0340(8)$ Ha/electron for the connectivity scheme and $-0.033(4)$ Ha/electron for the twist-averaged scheme, a difference of $0.001(4)$ Ha/electron. The numbers in parentheses are errors in the final digit. These agree within error, and the connectivity scheme has an improved error due to having more data points.

{Next, we demonstrate how to use this method to obtain a complete basis set and thermodynamic limit estimate for the uniform electron gas. 
Connectivity scheme CCD energies were collected for the $N=54$ electron system with basis sets varying from $M=922$ to $M=2838$ orbitals, and for systems with electron numbers varying between $N=162$ to $610$, with $M\approx 4N$. 
These data allow us to extrapolate to both the complete basis set limit and the thermodynamic limit by using the numerical approach set out in our previous work.\cite{shepherd_communication:_2016}
This yields an energy that is 0.0566(6), with the error in parentheses resulting from the extrapolations; this is in good agreement with our prior estimate with significantly less error.\cite{shepherd_communication:_2016}
For more details the reader is referred to the Supplementary Information. }

\begin{figure}
\includegraphics[width=0.4\textwidth,height=\textheight,keepaspectratio]{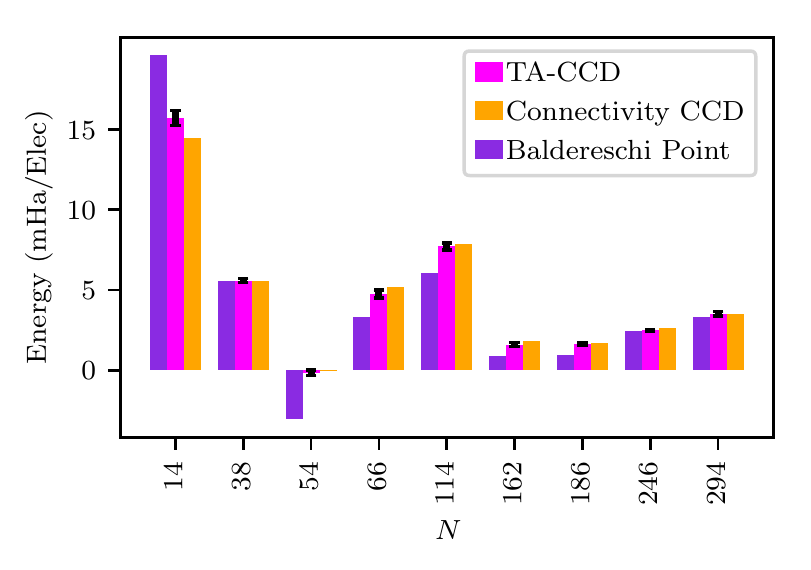}
\label{subfig:BPComp}

\caption{{All energies shown reflect the difference in correlation energy between the $\Gamma$-point and the relevant calculation.  The connectivity algorithm delivers comparable corrections to the correlation energy (relative to the $\Gamma$-point) when compared with twist averaging across a wide range of electron numbers. The Baldereschi point only delivers comparable corrections to the correlation energy (relative to the $\Gamma$-point) at higher electron numbers ($N \geq 162$) when compared with twist averaging. Twist averaging is performed over 100 twist angles. Standard errors are calculated in the normal fashion for twist averaging, $\sigma\approx\sqrt{\mathrm{Var}(E_{\mathrm{CCD}}({{\bf k}_s})) / N_s}$.}}
\label{fig:BPCompresults}
\end{figure}

{ Finally, in \reffig{fig:BPCompresults}, we compare the CCD energies from full twist-averaging, our connectivity scheme, and performing single calculation using the Baldereschi point as a twist angle. 
This point, first developed for insulators, is well known for the role it played in developing efficient thermodynamic integrations\cite{baldereschi_mean-value_1973,chadi_special_1973,cunningham_special_1974,monkhorst_special_1976} and was subsequently used for twist-averaging as the center-point of uniform grid twist-averaging by Drummond \emph{et al.}\cite{drummond_quantum_2009}.
At higher electron numbers ($N\geq 162$) the difference between BP and the TA energies falls below 1mHa/electron as all of the approaches converge to the same energy. 
At small electron numbers, however, the Baldereschi point significantly deviates from the twist-averaged energy, while the connectivity scheme is a much better approximation.}

\section{Discussion \& concluding remarks}

Our results show that a finite electron gas is best able to reproduce the twist-averaged total and correlation energies when a special $\bf{k}_s$-point is chosen to minimize the differences between the momentum connectivity of the finite system and a reference (here, a twist-averaged finite system). 
{Our interpretation of the connectivity-derived special $\bf{k}_s$-point's utility is that the low-momentum two-particle excitations from HF often suffer from finite size errors due to the shape of the Fermi surface in $k$-space. By finding a particularly representative $k_s$-point, we aim to take the `best case' of a representative shape--or, at least, as best as can be managed by a truly finite system. }
When we examine the occupied orbitals in $k$-space at the special $\bf{k}_s$-point, they adopt low-symmetry patterns that tend more toward the shape of a sphere than the $\Gamma$-point distribution.

Though we have made significant progress here towards ameliorating finite size error, {there are still two open questions. First, could our method be modified in order to minimize the energy difference to the thermodynamic limit rather than just to the twist-averaged energy? 
The second open question surrounds the extrapolation -- 
in particular, what is the \emph{actual} form of the energy as the system size tends to infinity?}
We could investigate this source of error by comparing with the known high-density limit of RPA, which CCD is expected to be able to capture. 
We leave both of these investigations for future work. 

Overall, the results here should improve our ability to understand infinite-sized model systems that are necessarily represented as finite systems, such as the electron gas with varying dimensions, the Hubbard model, and the models of nuclear matter we previously studied. \cite{shepherd_communication:_2016,Baardsen2,Baardsen1}
This communication is timely due to a resurgence of interest in the uniform electron gas~\cite{neufeld_study_2017,white_time-dependent_2018,spencer_large_2018,mcclain_spectral_2016,spencer_hande-qmc_2019,malone_accurate_2016,shepherd_many-body_2013,shepherd_range-separated_2014,gruneis_explicitly_2013}
 and of twist-averaged coupled cluster calculations.~\cite{gruber_applying_2018,hagen_coupledcluster_2014} We expect this work can immediately be applied to improve calculations.

{
Our long-term goals are to use this approach to study realistic systems. Though calculations are left for future manuscripts, we expect to follow a similar approach to our prior work in this area. 
In particular, we start by observing the similarity between how twist-averaging works in plane wave ab initio calculations where the energy is still obtained as a sum over matrix elements $v_{ijab}$ (as in \refeq{eq:ERIs}) which are offset by a twist angle.
Specifically, then, it should be possible to choose the twist angle in the same way as we propose here, so for a cubic system with $N$ electrons and a box length of $L$, the same twist angle as used here should work.
As such, we will soon be applying this to real solids and leave this for a future study.
}

{\bf \emph{Supplementary Material.--} } The reader is directed to the supplementary material for raw data tables and illustrations mentioned in the text. 

{\bf \emph{Acknowledgements.--} } JJS and TM acknowledge the University of Iowa for funding. JJS thanks the University of Iowa for an Old Gold Award. ARM was supported by the National Science Foundation Graduate Research
Fellowship under Grant No. 1122374. The code used throughout this work is a locally modified version of a github repository used in previous work~\cite{shepherd_range-separated_2014,shepherd_coupled_2014}: https://github.com/jamesjshepherd/uegccd.

%
%
%
%
%
%
%
%
%


%

\end{document}